\documentclass[epj,draft]{svjour}
\begin{document}
\input epsf
\def\paper#1#2#3#4#5{#1, #2 {\bf #3} (#5) \rm #4}
\def\cn{\mathop{\rm cn}\nolimits}
\def\dn{\mathop{\rm dn}\nolimits}
\def\sn{\mathop{\rm sn}\nolimits}
\def\be{\begin{equation}}
\def\ee{\end{equation}}
\def\bsigma{\mbox{\boldmath $\sigma$}}
\def\uhp{{\sc uhp}}
\def\fbc{{\rm f\sc bc}}
\def\Fbc{{\sc Fbc}} 
\def\bsigma{\mbox{\boldmath $\sigma$}}
\def\bmu{\mbox{\boldmath 
     $\mu$}}
\def\smallbmu{\mbox{\tiny\boldmath 
     $\mu$}}
\def\be{\begin{equation}}
\def\ee{\end{equation}}

\title{Influence of quenched dilution on the quasi-long-range ordered
  phase of the $2d$ $XY$ model.}

\titlerunning{Influence of dilution on the $2d$ $XY$ model}

\author{B. Berche\inst 1, A.I. Fari\~nas-S\'anchez\inst{1,2},
    Yu. Holovatch\inst{3,4} and
  R. Paredes V.\inst 2}

\authorrunning{B. Berche, A.I. Fari\~nas-S\'anchez,
    Yu. Holovatch and R. Paredes V.}

\institute{Laboratoire de Physique des Mat\'eriaux, 
    Universit\'e Henri Poincar\'e,  Nancy 1,\\
    F-54506 Vand\oe uvre les Nancy Cedex, France
    \and
    Centro de F\'\i sica, Instituto Venezolano de 
    Investigaciones Cient\'\i ficas,\\
    Apartado 21827, Caracas 1020A, Venezuela
    \and
	Institute for Condensed Matter Physics,
	National Academy of Sciences of Ukraine, \\
	1 Svientsitskii Str., Lviv, 79011 Ukraine 
	\and
	Ivan Franko National University of Lviv, 
	12 Drahomanov Str., Lviv, 79005 Ukraine
}

\date{\today}

\abstract{
    The influence of non magnetic impurities in the $2d$ $XY$ model
    is investigated through Monte Carlo (MC) simulations. The general picture
    of the transition is fully understood from the Harris criterion which
    predicts that the universality class is unchanged, and the 
    Berezinskii-Kosterlitz-Thouless description of the topological
    transition remains valid. 
We nevertheless address here the question about the influence of
dilution on the quasi-long-range order at low temperatures. In particular,
we study the asymptotic of the pair correlation function and report the
MC estimates for the critical exponent $\eta$ at different dilutions.
In the weak dilution region, our MC calculations are further supported
by simple spin-wave-like calculations.
  }

\maketitle

\section{Introduction}\label{Intro}
From general scaling arguments, Harris criterion predicts that Gaussian
disorder (coupled to the energy density)  is an irrelevant perturbation 
at the fixed point of any pure
system when the exponent $\alpha$ which governs the singular behaviour
of the specific heat of the pure model in the vicinity of the critical 
point is negative~\cite{Harris74}.
The critical behaviour of the random system is thus governed by the pure
fixed point and the universality class of the model remains unchanged.

From this respect, the $2d$ $XY$ model is probably the one where the influence
of quenched disorder is essentially trivial, and nothing special is expected,
since it displays a critical behaviour described by essential singularities.
On the other hand, the transition of the $2d$ $XY$ model is not conventional
and has very interesting features which make the question of the influence
of disorder worth studying. 
Indeed, the transition, first described by 
Berezinskii~\cite{Berezinskii71} and Kosterlitz and 
Thouless~\cite{KosterlitzThouless73,Kosterlitz74}, is governed by 
the condensation of topological defects.
The symmetry of the $XY$ model being continuous, there is 
no spontaneous order in the system (in $2d$)
at any non zero temperature according to
Mermin Wagner theorem~\cite{MerminWagner66,Hohenberg67}. Nevertheless 
at very {\it low 
temperatures}, 
the spin-wave approximation which assumes that the disorientation
$\theta_{\bf r}-\theta_{{\bf r}+{\bmu}}$ 
between neighbouring spins ($\bmu$
denotes the unit lattice spacing) remains 
small, captures the essential features
of the behaviour of the system, leading to an algebraic decay of the 
correlation function~\cite{Wegner67}
\be
\langle\bsigma({\bf r})\cdot\bsigma({\bf r+R})\rangle\sim|{\bf R}|^{-\eta(T)},
\label{eq-1}
\ee
(where $\bsigma({\bf r})$ are two-component unit vectors), 
and, as a consequence, 
to the absence of magnetization and to an infinite susceptibility. The
correlation function exponent, $\eta(T)$, continuously increases with 
temperature.
The model is said to have a low temperature phase with {\it quasi-long-range}
order. At {\it high temperatures} on the other hand, a high temperature series
expansion leads to a more conventional 
exponential decay of the correlation function. 
It is clear from these two extreme behaviours that something must happen 
in between and that the model undergoes a transition in the intermediate 
regime. The scenario proposed by Kosterlitz and Thouless (KT) is based on the
existence of vortices, which are localized defects where the 
field $\theta_{\bf r}$ (the angle between spin $\bsigma({\bf r})$ and some
arbitrary reference direction) may become singular at some points. 
The energy carried by
such defects increases logarithmically with the system size, and thus they
are constrained to be associated in pairs which nevertheless
amplify the disordering of 
the system, leading to an effective increase of the temperature. 
The vortices appear in increasing number in the system when the temperature
increases, and the transition temperature is reached when the pairs
break, leaving the system totally disordered. In the high temperature
phase, the approach to the critical point $T_{c}$, (also referred to as
KT temperature, $T_{\rm KT}$), 
is described by essential 
singularities, for instance, the correlation length and the 
susceptibility behave
as  $\xi(t)\sim\exp (-bt^{-\sigma})$,
$\chi(t)\sim\xi^{2-\eta}$, $t=|T-T_{\rm KT}|$, with $\eta=\frac 14$ and 
$\sigma=\frac 12$
(for reviews, see 
e.g.~\cite{KosterlitzThouless78,Nelson83,ItzyksonDrouffe89,GulasciGulasci98,ChaikinLubensky95}).

From the hyperscaling relation $\alpha=2-\nu d$, and
from the fact that the correlation length has an essential singularity
at the critical point, one deduces a non typical value
for the exponent of the specific heat, $\alpha=-\infty$, which implies, 
according to Harris criterion, that the adjunction of thermal randomness
(i.e. coupled to the energy density) does not modify the general
scenario described above. All along the transition
line, the KT universality class might be recovered (that is an exponent
$\eta(T_{\rm KT})=\frac 14$ for the correlation function decay).
Let us specify the case of site dilution for example. The original Hamiltonian
of the pure $2d$ $XY$ model,
\be
H_0=-J\sum_{{\bf r}}\sum_{\smallbmu}
\bsigma({\bf r})\cdot \bsigma({\bf r}+\bmu),
\label{eq1}
\ee
where $\bsigma({\bf r})\cdot \bsigma({\bf r}+\bmu)$ denotes the scalar product
and the spins $\bsigma({\bf r})$ are located at the lattice 
sites ${\bf r}$ of a square lattice $\Lambda$, 
is modified by the introduction of
a set of occupation variables $c_{\bf r}$ which take the values 0 (with
probability $1-p$) or
1 (with probability $p$)
depending on the fact that site ${\bf r}$ is empty or occupied by a
spin,
\begin{eqnarray}
{\cal P}[c_{\bf r}]&=&\prod_{{\bf r}}P (c_{\bf r})\nonumber\\
&=&\prod_{{\bf r}}[p\delta(c_{\bf r}-1)+
(1-p)\delta(c_{\bf r})].
\label{eq2}
\end{eqnarray}
Thus the Hamiltonian of the diluted $2d$ $XY$ model reads:
\be
H=-J\sum_{{\bf r}}\sum_{\smallbmu}
c_{\bf r}c_{{\bf r}+\smallbmu}\ \!\bsigma({\bf r})\cdot \bsigma({\bf r}+\bmu).
\label{eq3}
\ee
Altogether, one expects a phase diagram starting from the
pure system critical temperature 
$k_BT_{\rm KT}/J\simeq 0.893$~\cite{HasenbuchPinn97} at $p=1$, and 
decreasing up to a transition at zero temperature at the site percolation
threshold of the $2d$ square lattice, $p_c\simeq 0.59$,
since no transition at all takes place in the system when there
is no more percolating cluster of spins.  
The question which remains interesting is to understand the exact role of the
impurities. Of course, the adjunction of impurities would first decrease the
transition temperature through the usual dilution effect which, at a
mean field approximation decreases the average coordination number.
But in the same time, the number of vortices (and thus also their 
disordering consequences) will possibly decrease.
Also their interactions between
each other and the interactions between vortices and impurities might
play some role. This question has been partly discussed in 
refs.~\cite{MolPereiraPires02,LeonelEtAl02} and we essentially
address here the question of
the role of impurities at low temperature, where the spin-wave approximation
should give reliable results.

\section{Determination of the phase diagram of the diluted model}\label{sec1}
The determination of the phase diagram is performed using a fit of the
order parameter profile inside a finite system to the functional expression
predicted by a convenient conformal mapping,
valid at a scale-invariant critical point. By extension it is also valid
in the {\it whole low-temperature phase} of the $XY$ model which displays 
scale-invariant algebraic correlation functions.
This method has been applied with success to the case of the pure $XY$
model~\cite{ResStraley00,BercheEtAl02,Berche03} and should provide 
here also reliable
results. The order parameter vanishes in the bulk 
of the system at any 
temperature in the $XY$ model, unless symmetry breaking fields are applied
along some boundaries $\partial\Lambda$ for example. The magnetization profile 
$\langle\bsigma({\bf r})\cdot\bsigma_{\partial\Lambda}\rangle$ thus
obeys a general covariance law under conformal transformations. The
case of a square geometry with fixed spins along its fours edges is 
particularly easy to implement in Monte Carlo simulations. There, the effect
of the Schwarz-Christoffel conformal mapping is just to define a rescaled
distance variable, called $\kappa(w)$ (here and in the following, $w$
stands for the complex variable associated to the point ${\bf r}$), in terms
of which one recovers inside the square with fixed boundary conditions, a
simple power law for the profile:
\be
\langle\bsigma(w)\cdot\bsigma_{\partial\Lambda}\rangle
\sim[\kappa(w)]^{-\frac 12\eta},\label{eq4}
\ee
with
\be
\kappa(w)= {\rm Im}\left[{\rm sn}\frac{2{\rm K}w}{L} \right]\times
        \left| 
	\cn\frac{2{\rm K}w}{L}\dn\frac{2{\rm K}w}{L}
	\right|^{-1/2}.
\label{eqkappa}
\ee
Here, $\cn x$, $\dn x$ and $\sn x$ are the Jacobi elliptic 
functions~\cite{Abramovitz64}, 
$L$ the linear size of the lattice 
$\Lambda$ and ${\rm K}\equiv {\rm K}(k)$ 
(the complete elliptic integral of the first kind)
and $k$ are constants related to the aspect ratio 
of the system. For more details, the reader is referred to 
ref.~\cite{Berche03}. 
The main advantage of this technique is that one lattice size is in principle 
sufficient (provided it is large enough), since the shape effects are 
included in the conformal mapping and
the method is not much sensitive to finite-size effects. The effect of
discretization of the lattice is only apparent at the scale of a few 
lattice spacings. One more advantage is the fact that all the information
encoded in the profile is used, since all the points $w$ inside the square
are taken into account in the fit. 
The strategy to obtain the phase diagram  is summarised  in the following: 

i) Perform simulations on a small system
($32\times  32$) for several dilutions (from $p=0.50$ to $1.00$ every 
$0.05$) and several temperatures (from $k_BT/J=0.05$ to 1.25 every 0.05).

ii) Fit the data to equation~(\ref{eq4}) (from now on, we use
the notation $m(w)$ for 
$\langle\bsigma(w)\cdot\bsigma_{\partial\Lambda}\rangle$), 
compute from a least-square fit the sum of deviation squares, $\chi^2$, per
degree of freedom (d.o.f.). 

iii) Plot the chi-square per degree of freedom, $\chi^2/d.o.f.$, 
as a function of temperature
for all dilutions. It should keep a small value (the better the disorder
statistics the smaller the expected value) in the scale-invariant phase
and then increase sharply in the paramagnetic phase where 
expression~(\ref{eq4}) is no longer valid. 

iv) Report on the phase diagram the temperature where the $\chi^2/d.o.f.$ 
starts to increase
sharply, and which is the estimate of the critical temperature at 
a given dilution.
The results will be refined later when larger lattices will be considered
in order to investigate the universality class at the transition.

We may also mention here a few comments about the technical details: 
we use a Wolff 
algorithm~\cite{Wolff89} and each sample is thermalized
by $10^4$ Wolff iterations, and $10^4$ other sweeps are used for computation
of the physical quantities. 
The average over disorder realizations is 
performed over $10^4$ samples. One iteration takes of the order of 1~$\mu$s 
of CPU time per spin on a standard processor, so one simulation needs around
$200\times L^2$ seconds (and here we have 23 temperatures times 10 dilutions).
For more precise estimates to be described later, we also performed
simulations for sizes $64\times 64$ (with $4\cdot 10^4$ samples for disorder 
average, so one simulation takes $8000\times L^2$ seconds) and 
$128\times 128$ (with $2\cdot 10^4$ samples for disorder 
average, or $4000\times L^2$ seconds for one simulation).  

\begin{figure} [ht]
\vspace{0.2cm}
        \epsfxsize=8.5cm
        \begin{center}
        \mbox{\epsfbox{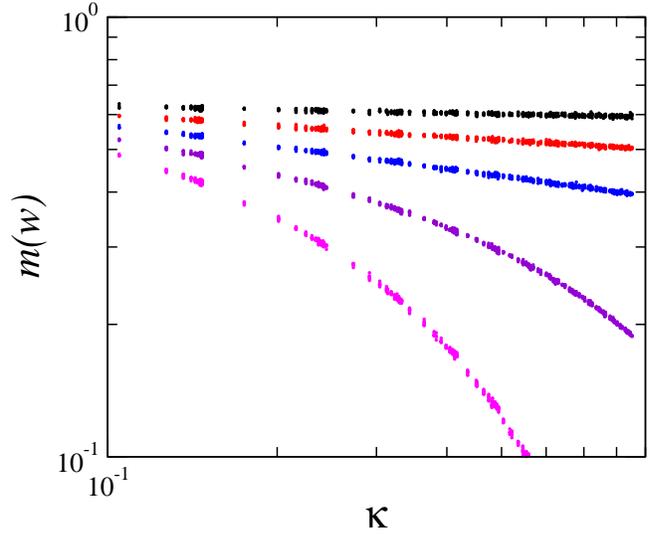}\qquad}
        \end{center}
        \caption{Order parameter profile for several temperatures 
          ($k_BT/J=0.1$, 0.3; 0.5; 0.7 and 0.9 from top to bottom) at a 
          dilution $p=0.80$ as a function of the rescaled variable 
          $\kappa(w)$. The transition is below $k_BT/J=0.5$.}
        \label{fig1}  \vskip -0cm
\end{figure}

In order to illustrate the above mentioned programme, we first show in
figure~\ref{fig1} the
behaviour of the order parameter profile for several temperatures at a 
dilution $p=0.80$ as a function of the rescaled variable $\kappa(w)$ on
a log-log scale.
It is particularly clear that the data fit quite nicely to a power law
at low temperatures while this type of fit completely breaks down at higher
temperatures.

\begin{figure} [ht]
\vspace{0.2cm}
        \epsfysize=5cm
        \begin{center}
        \mbox{\epsfbox{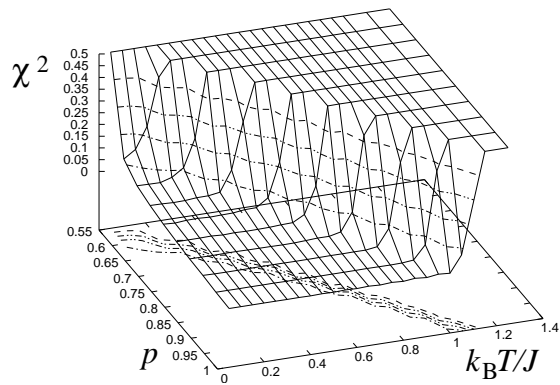}\qquad}
        \end{center}
        \caption{$3d$ plot of the $\chi^2$ (to get it per d.o.f., one has to
          divide by $L^2$)
          as a function of 
          temperature and dilution for a system of size $32\times 32$.}
        \label{fig2}  \vskip -0cm
\end{figure}

\begin{figure} [ht]
\vspace{0.2cm}
        \epsfxsize=8.5cm
        \begin{center}
        \mbox{\epsfbox{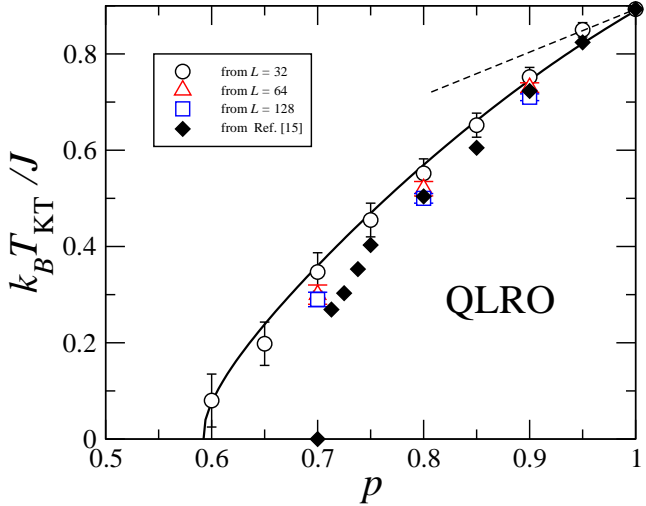}\qquad}
        \end{center}
        \caption{Phase diagram of the $2d$ dilute $XY$ model. The data points
        correspond to the results obtained after fitting the order parameter
        to the conformal expression. They correspond to the temperature where
        the $\chi^2$ has a sudden increase. The error bars are estimated 
        empirically. Three different sizes are represented and the solid line 
        is only a guide for the eyes. The full symbols also reported were taken
        from the work of Leonel et al.~\cite{LeonelEtAl02}.
	Dashed line shows the mean field prediction.
}
        \label{fig3}  \vskip -0cm
\end{figure}

A $3d$ plot
of the $\chi^2$ in the temperature-dilution plane
is then presented in figure~\ref{fig2}. For the sake of clarity, a cutoff
at 0.5 was introduced in order to avoid too large $\chi^2$ at high 
temperatures. The low temperature phase with quasi-long-range order extends in 
the whole region in the $(p,T)$ plane where the $\chi^2$ is close to zero,
revealing that equation~(\ref{eq4}) nicely fits the numerical data.
The phase diagram which can be deduced from these data is shown in
figure~\ref{fig3} where the data from larger system sizes are also reported,
as well as previous results~\cite{LeonelEtAl02}. 
As expected, as a result of the influence of dilution, 
the transition temperature decreases from the pure system value at $p=1$
and it vanishes at the percolation threshold
of site percolation on the square lattice.

A simple standard mean field argument for dilution in the low impurity
concentration regime ($p$ close to 1) gives the beginning of the transition 
line in the vicinity of the pure system. The coordination number $z$ on the
lattice with zero impurity concentration becomes $pz$ when only a (small) 
fraction $1-p$ of sites is unoccupied 
(this is of course correct only to leading 
order in the very neighbourhood of $p=1$, since the problem under interest here
is {\it site dilution} and not {\it bond dilution}) 
and thus the transition temperature 
may be estimated by
${k_BT_{\rm KT}(p)}/{J}=p{k_BT_{\rm KT}(p=1)}/{J}\simeq 0.893p$.
This transition curve is shown in dashed line in figure~\ref{fig3} and fits 
correctly only the first point at $p=0.95$.

In the vicinity of the percolation threshold, one would expect that the 
transition temperature increases with the fraction of sites belonging to the 
percolating cluster. From this argument a power-law behaviour would follow,
$T_c\sim (p-p_c)^{5/36}$, with the exponent of the order parameter in $2d$
percolation (see for example
Refs.~\cite{KatoEtAl00,Sandvik02}). We cannot check this behaviour in the
present
model where we essentially focus on smaller dilutions ($p$ closer to 1) and
we did not perform enough simulations close to the percolation threshold.


\section{Low temperature behaviour of the diluted $XY$ model}\label{sec2}
In this section, we address the question of the behaviour of the diluted $XY$
model at low temperatures and of the universality class along the transition
line. More precisely, does the spin-wave description provide the correct
behaviour at low temperatures in the presence of dilution, and does one
recover the Kosterlitz-Thouless universality class at the transition 
temperature between the phase with quasi-long-range order and the paramagnetic
phase?

First of all, in order to bring reliable arguments, we have to 
produce more precise data at larger sizes
to approach the thermodynamic limit. 
We performed simulations 
for systems of sizes
$L=64$ and 128, but only for a few temperatures in the vicinity of the
transition (estimated from the previous results for the system of size $L=32$)
and for three dilutions, $p=0.70$, 0.80 and 0.90. The data are
shown in figure~4. On the left we plot the $\eta$ exponent
obtained from a least square fit of $\ln m(w)$ vs $\ln \kappa(w)$ 
and on the right, we
show the corresponding $\chi^2/d.o.f.$. The vertical stripe 
in each figure is a rough
estimate of the transition temperature, where $\eta$ takes its KT value
$\frac 14$ and $\chi^2/d.o.f.$ is at the edge of the plateau region (which
means that order parameter density profiles
are still fitted by power laws, i.e. the system is still critical). 
The larger the system size, the smaller the value of $T$ where the
$\chi^2/d.o.f.$ increases and the sharper the increase (note the 
logarithmic scale on the vertical axis). 
We also notice 
that the absolute value of the $\chi^2/d.o.f.$ in the quasi-long-range 
ordered phase is meaningless, since it is strongly dependent on the number of
disorder realisations. Here the smaller value is obtained for $L=64$ where
40000 samples were produced while only 20000 were realized at $L=128$.
The horizontal scale has
been chosen identical for all three figures, since it facilitates the 
comparison and makes obvious the role of dilution.

Let us now discuss the low temperature limit. There, a spin-wave 
calculation (given in the appendix),
based on the assumption that the spin disorientation remains small and 
allows to expand the $\cos$ in the definition of the Hamiltonian,
shows that the correlation function exponent is simply modified by a 
``rescaling'' of temperature due to the presence of the impurities.
While the pure model exponent is given by
\begin{equation}\label{27last}
 \eta^{\rm pure}_{SW}=\frac{k_BT}{2\pi J},
 \end{equation}
we get in the case of the disordered system
\begin{equation}\label{28}
 \eta^{\rm diluted}_{SW}=\frac{k_BT}{2\pi J(1-2(1-p))}.
 \end{equation}
Within the spin-wave approximation, the $\eta$-exponent still varies
linearly with temperature at low temperatures, but as expected,
it increases faster
in the disordered model. The role of impurities is simply described by
an augmented effective temperature.
We also note that
the value of the coefficient, $k_B/2\pi J(1-2(1-p))$,
is not simply obtained from a replacement of the nearest-neighbour
coupling $J$ by its average $pJ$ in the low dilution limit.

\begin{table}
\begin{center}
\begin{tabular}{lcccc}
\hline
$p$ &  $k_BT/J$ & $\eta_{\rm MC}$ & $\eta_{\rm SW}^{\rm diluted}$
& $\eta_{\rm SW}^{\rm pure}$ \\
\hline
0.96 & 0.04 &  0.007(2) &  0.0069  & 0.0064 \\
$--$ & 0.08 &  0.016(2) &  0.0138  & 0.0127 \\
$--$ & 0.12 &  0.024(2) &  0.0208  & 0.0191 \\
$--$ & 0.16 &  0.032(2) &  0.0277  & 0.0255 \\
\hline
0.92 & 0.04 &  0.009(2) &  0.0076  & 0.0064 \\
$--$ & 0.08 &  0.018(2) &  0.0152  & 0.0127 \\
$--$ & 0.12 &  0.027(2) &  0.0227  & 0.0191 \\
$--$ & 0.16 &  0.036(2) &  0.0303  & 0.0255 \\
\hline
0.88 & 0.04 &  0.010(2) &  0.0084  & 0.0064 \\
$--$ & 0.08 &  0.021(2) &  0.0168  & 0.0127 \\
$--$ & 0.12 &  0.032(2) &  0.0251  & 0.0191 \\
$--$ & 0.16 &  0.043(2) &  0.0355  & 0.0255 \\
\hline
\end{tabular}
\end{center}
\caption{Comparison between the Monte Carlo results and the spin-wave
	approximation for the $2d$ dilute $XY$ model. The numerical data were 
	obtained by fitting the density profile inside a square of size
	$L=128$. For comparison, the results of the spin-wave approximation
	of the corresponding pure system are also presented.}
\label{tab1}
\end{table}

A comparison with numerical results (in table~\ref{tab1}) 
deduced from Monte Carlo simulations
shows that the spin-wave approximation fits correctly the numerical results
at very low temperature and low empty sites concentration. It is worth noticing
that the numerical data were obtained on a finite system, and that increasing
the system size still decreases slightly 
the resulting exponent (see fig.~4).

\onecolumn

\vspace{0.2cm}
\begin{figure}
        \begin{center}
        \epsfysize=6.4cm
        \mbox{\epsfbox{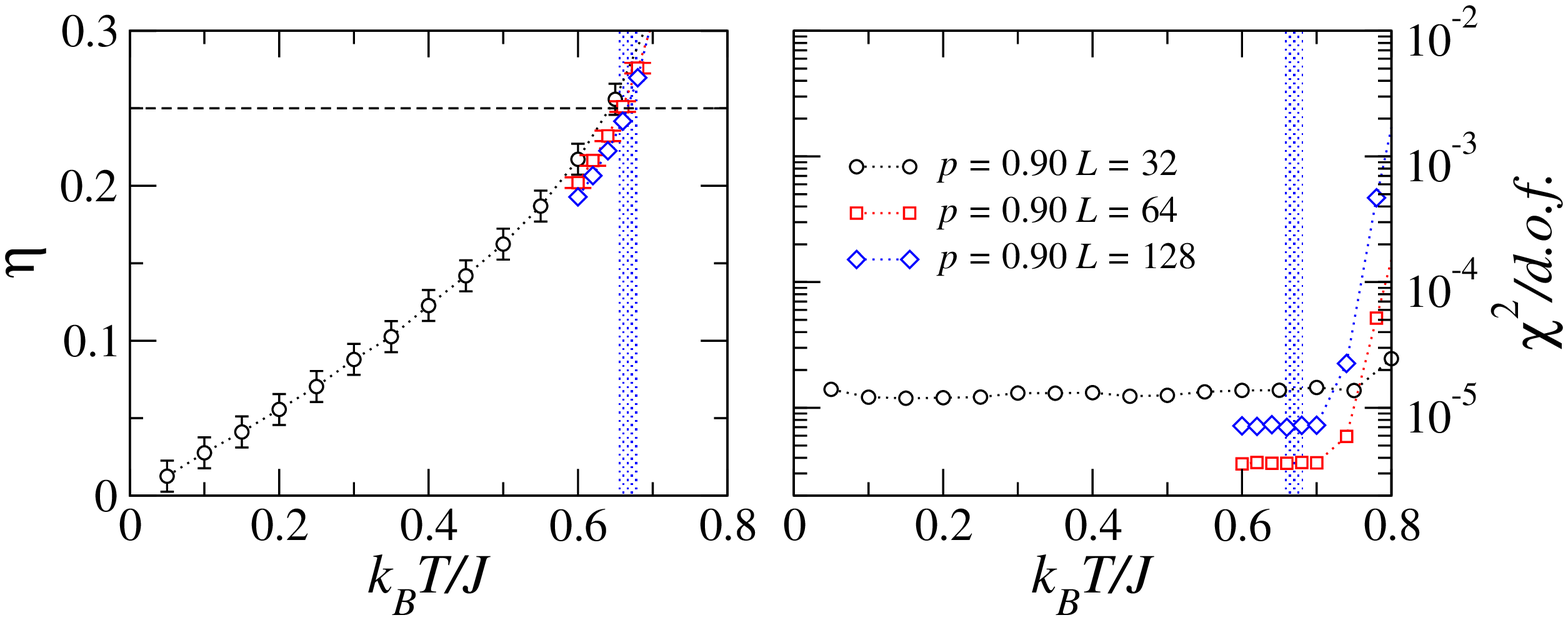}\qquad}
        \epsfysize=6.4cm
        \mbox{\epsfbox{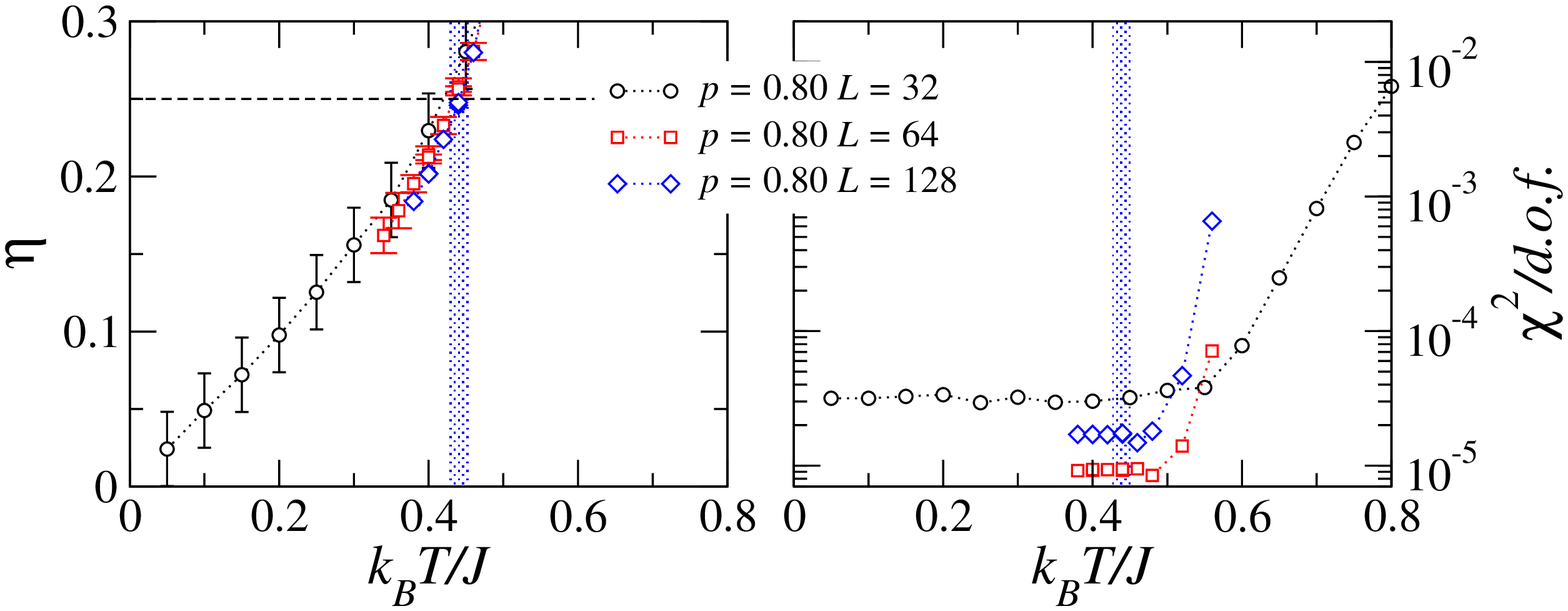}\qquad}
        \epsfysize=6.4cm
        \mbox{\epsfbox{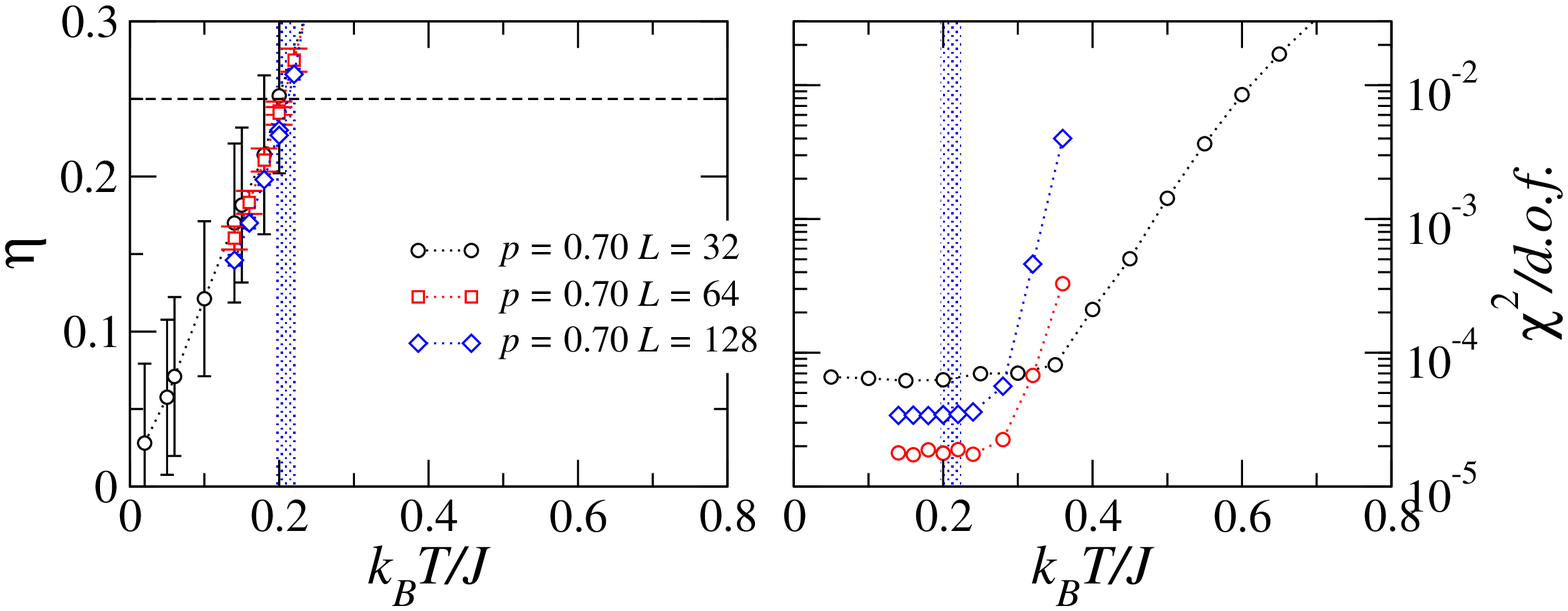}\qquad}
        \end{center}
	\caption{Plot of the exponent $\eta$ as a 
          function of the temperature (left)
          for three different dilutions. The figures on the right give the
          $\chi^2/d.o.f.$ corresponding to the fit of the order parameter
          profiles. The vertical stripe materialises the transition temperature
          where the $\chi^2/d.o.f.$ suddenly increases (logarithmic scale).
	}
	\label{fig4}  \vskip -0cm
\end{figure}

\twocolumn

\section{Conclusions}
We have performed an extensive Monte Carlo study of the critical behaviour
in the quasi-long-range ordered phase of the two-dimensional diluted
$XY$ model. According to Harris criterion, one expects weak disorder
to be irrelevant at the Kosterlitz-Thouless transition where the pure
model exhibits essential singularities. The numerical results confirm this
picture, since
at the transition temperature, we observe from figure~4 that the
$\eta$ exponent is compatible with the constant value $\frac 14$. 

\begin{figure} [h]
\vspace{0.2cm}
        \epsfysize=7.cm
        \begin{center}
        \mbox{\epsfbox{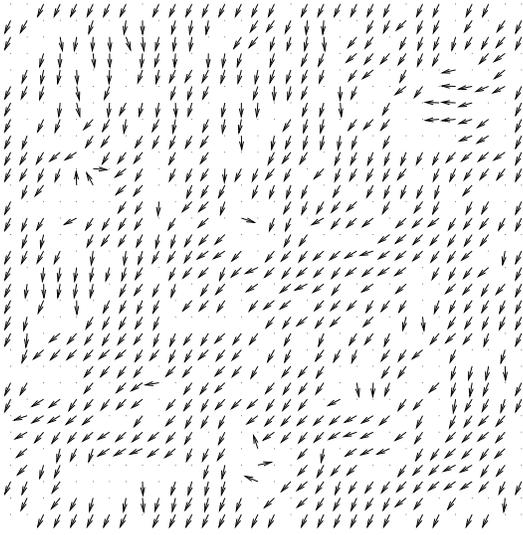}\qquad}
        \end{center}
        \caption{Typical configuration for a system of size 
	$32\times 32$, $p=0.70$ at low temperature $k_BT/J=0.05$.}
        \label{fig0}  \vskip -0cm
\end{figure}

We also obtained the approximate spin-wave solution which describes the
properties of the diluted model at very low temperatures. The presence of
non magnetic impurities produces an effective increase of the temperature, 
but does not completely suppress the transition before the percolation
threshold is reached, as it was suspected in ref.~\cite{LeonelEtAl02} where a
zero-temperature transition was reported at $p\simeq 0.70$.
The argument reported in ref.~\cite{LeonelEtAl02} was that impurities, when
located in the vicinity of a pair, produce
a repulsive interaction between vortices
which facilitates pair unbinding. This scenario, together with dilution effect,
enhances the disordering of the system but is not sufficient to prevent
quasi-long-range ordering at impurity concentrations above the percolation 
threshold. As an example, a typical configuration at low temperature 
is shown in fig.~\ref{fig0} for a system of size 
$32\times 32$ at $p=0.70$.

In conclusion, one should stress that the behaviour of the model in the
low temperature critical phase is in full agreement with what might be
expected from general relevance arguments.

\begin{acknowledgement}
{\bf Acknowledgement:} This work is supported by the French-Venezuelan PCP
program  `Fluides p\'etroliers' and by the French-Ukrainian cooperation
Dnipro project. The authors gratefully acknowledge both programs for their
support. Support from the CINES under project c20020622309 is also
gratefully acknowledged. B.B. warmly thanks C. Chatelain for his 
friendly technical cooperation and Yu.H. acknowledges useful discussions with 
V. Tkachuk.
\end{acknowledgement}

\begin{appendix}
\section*{Appendix: spin-wave calculation at low temperature}
\label{appendix}
In this appendix, we present the result of a spin-wave calculation valid
in the low temperature limit.  
First, we rewrite
the Hamiltonian (\ref{eq3}) of the diluted $2d$ $XY$ model for
the case of arbitrary short-range ferromagnetic interaction
potential $J(|{\bf r}|)$:
\begin{eqnarray}\label{1}
{H}
&=&
-\frac{1}{2}\sum_{{\bf r}\neq {\bf r}^{\prime}}
J(| {\bf r}-{\bf r}^{\prime}|)
\cos(\theta_{\bf r}
-\theta_{{\bf r}^{\prime}}) c_{\bf r} c_{{\bf r}^{\prime}}.
\end{eqnarray}
Here, $\cos(\theta_{\bf r}-\theta_{{\bf r}^{\prime}})$ stands for
a scalar product of two-component unit vectors
\bsigma({\bf r}) directed by angles $\theta_{\bf r}$,
$\theta_{{\bf r}^{\prime}}$ (c.f. Eq. (\ref{eq1})).
Expanding $\cos(\theta_{\bf r} -\theta_{{\bf r}^{\prime}})$ in (\ref{1}) for
small difference in directions of  spins one gets
the Hamiltonian (\ref{1}) in the spin-wave approximation:
\begin{eqnarray}\label{2}
{H}&\simeq& {H}'+\frac{1}{4}\sum_{{\bf r}\neq
{\bf r}^{\prime}} J(| {\bf r}-{\bf r}^{\prime}|) (\theta_{\bf r} -\theta_{{\bf r}^{\prime}})^2
c_{\bf r} c_{{\bf r}^{\prime}}\nonumber\\
&\equiv& {H}' + {H}_1.
\end{eqnarray}
The term ${H}'$ in the Hamiltonian (\ref{2}) will
be further absorbed into the energy reference point, whereas to
rewrite the remaining part ${H}_1$ we pass to the
Fourier-trans\-formed angle variables and the interaction potential.
Taking the periodic boundary conditions we define the
Fourier-trans\-formed quantities as:
\begin{eqnarray}\label{3}
&& \theta_{\bf r}=\frac{1}{\sqrt{N}}\sum_{\bf k} 
e^{i{\bf k.r}}\theta_{\bf k}, \ 
\theta_{\bf k}=\frac{1}{\sqrt{N}}\sum_{\bf r} e^{-i{\bf k.r}}\theta_{\bf r}.
\\ \label{4} && J({\bf r})=\frac{1}{N}\sum_{\bf k} e^{i{\bf k.r}}
\nu ({\bf k}),\ 
\nu ({\bf k})=\sum_{\bf r} e^{-i{\bf k.r}}J({\bf r}).
\end{eqnarray}
In (\ref{3}), (\ref{4}) $\sum_{\bf r}$ spans $N$ sites of the lattice
whereas $\sum_{\bf k}$ is within the first Brillouin zone. Written in
the Fourier-transformed variables the Hamiltonian ${H}_1$
reads:
\begin{eqnarray}\label{5}
{H}_1=&& \frac{1}{2}\sum_{{\bf q},{\bf k}_1,{\bf k}_2} 
S({\bf q}+{\bf k}_1+{\bf k}_2) S(-{\bf q}) \nu({\bf q})
\theta_{{\bf k}_1} \theta_{{\bf k}_2}\nonumber\\
 &-& \frac{1}{2}\sum_{{\bf q},{\bf k}_1,{\bf k}_2} S({\bf q}+{\bf k}_1)
S({\bf k}_2-{\bf q}) \nu({\bf q}) \theta_{{\bf k}_1} \theta_{{\bf k}_2},
\end{eqnarray}
with
\begin{equation}\label{6}
S({\bf q})= \frac{1}{N}\sum_{{\bf r}} e^{i{\bf q.r}}c_{\bf r}.
\end{equation}

Let us single out in (\ref{5}) the Hamiltonian of the pure $2d$ $XY$
model. To this end let us introduce variables $\rho({\bf q})$ which may
serve to show deviation of $S({\bf q})$ from the Kroneker's delta
$\delta({\bf q})$:
\begin{equation} \label{7}
S({\bf q})=\delta({\bf q})-\rho({\bf q}),
\end{equation}
with
\begin{equation} \label{8}
\rho({\bf q})=\frac{1}{N}\sum_{{\bf r}} e^{i{\bf q.r}}(1-c_{\bf r}).
\end{equation}

Denoting a configurational average with the distribution function
(\ref{eq2}) by $\overline{(....)}$ we get for the two first moments
of the random variable $c_{\bf r}$:
$\overline{c_{\bf r}}=\sum_{c_{\bf r}=0,1}{P}(c_{\bf r})c_{\bf r}=p $
and $\overline{c_{\bf r}c_{{\bf r}^\prime}}= p^2$ for
$ {\bf r}\neq {\bf r}^\prime$, or 
 $p$ for $ {\bf r}={\bf r}^\prime$.
Here, we remember that $p$ 
is the concentration of the occupied sites $c_{\bf r}=1$ and that
${\cal P}[c_{\bf r}]=\prod_{\bf r}{P}(c_{\bf r})$.
This leads to the following values for the first two moments of
$\rho({\bf q})$ (note, that due to (\ref{1}) ${\bf r}\neq {\bf r}^\prime$):
 $\overline{\rho({\bf q})}= (1-p)\,\delta({\bf q})$ and 
 $\overline{\rho({\bf q})\rho({\bf q}^{\prime})}= (1-p)^2 \,\delta({\bf q})
 \delta({\bf q}^\prime)$.
Considering the concentration of empty sites ($1-p$) to be small
we take it as an expansion parameter and further we keep only
linear in $\rho({\bf q})$ contributions. For the Hamiltonian  (\ref{1})
we get:
\begin{eqnarray}\label{15}
{H}&=& {H}_0\nonumber
- \frac{1}{2}\sum_{{\bf k}_1,{\bf k}_2} \Big [
\nu({\bf k}_1+{\bf k}_2)
+ \nu(0) - \nu({\bf k}_1)\\ 
&&
- \nu({\bf k}_2) \Big ]
\rho({\bf k}_1+{\bf k}_2)\theta_{{\bf k}_1} \theta_{{\bf k}_2} + O(\rho^2).
\end{eqnarray}
Where ${H}_0$ stands for the Hamiltonian of the
undiluted system:
\begin{equation}\label{16}
{H}_0=  \frac{1}{2}\sum_{{\bf k}} \Big [ \nu(0) -
\nu({\bf k}) \Big ] \theta_{{\bf k}} \theta_{-{\bf k}}.
\end{equation}
For the nearest neighbour interaction and square lattice with a
lattice constant $a=|\bmu|$ we have
$J({\bf r})=  J$ for $|{\bf r}|=a$ and $0$ for
$|{\bf r}|\neq a$, 
\begin{equation} \label{17bis}
 \ 
\nu({\bf k})=2J(\cos(k_xa)+\cos(k_ya))\simeq 4J -
J|{\bf k}|^2a^2 + \dots
\end{equation}
Subsequently, the  Hamiltonian (\ref{15}) reads:
\begin{eqnarray}\label{18}
{H}&\simeq& \frac{1}{2}\sum_{{\bf k}} J|{\bf k}|^2a^2\theta_{\bf k} 
\theta_{-{\bf k}}\nonumber\\
&& +
\sum_{{\bf k}_1,{\bf k}_2} Ja^2({\bf k}_1.{\bf k}_2)\rho({\bf k}_1+{\bf k}_2)
\theta_{{\bf k}_1} \theta_{{\bf k}_2},
\end{eqnarray}
with a scalar product ${\bf k}_1.{\bf k}_2$.

For a given configuration of disorder, the
configura\-tio\-nal\-ly-dependent partition function ${Z}_{\rm
conf}$ is defined by:
\begin{equation}\label{19}
{Z}_{\rm conf} = {\rm Sp}_\theta e^{-\beta{H}},
\end{equation}
where ${\rm Sp}_\theta$ means integration over spin degrees of
freedom on each site (and $\beta=(k_BT)^{-1}$):
\begin{equation} \label{20}
{\rm Sp}_\theta (\dots)= \prod_{{\bf r}} \int_{-\pi}^{\pi}\frac{{\rm d}
\theta_{\bf r}}{2\pi} (\dots).
\end{equation}
For a given configuration of occupied and empty sites, let us
define a thermodynamic averaging by:
\begin{equation}\label{21}
\langle (\dots)\rangle =\frac{1}{{Z}_{\rm conf}} {\rm Sp}_\theta
e^{-\beta{H}}(\dots).
\end{equation}
Considering the quenched dilution (i.e. the case when magnetic and
non-magnetic sites are fixed on their places) one gets the
observables by averaging thermodynamically averaged
quantities with respect to different configurations of disorder
\cite{Brout59}.
Now, the pair correlation function  is defined as 
\begin{eqnarray}\label{22}
&&G_2(|{\bf r}_2-{\bf r}_1|)=\overline{\langle c_{{\bf r}_2}c_{{\bf r}_1}\cos(\theta_{{\bf r}_2}-\theta_{{\bf r}_1})\rangle }
\nonumber\\ \nonumber\\
&&\simeq \overline{ \frac{1}{{Z}_{\rm conf}} {\rm Sp}_\theta
e^{-\beta{H}} c_{{\bf r}_2}c_{{\bf r}_1}\Big (1 -
\frac 12(\theta_{{\bf r}_2}-\theta_{{\bf r}_1})^2\Big ) }.
\end{eqnarray}
As far as we keep only linear in $\rho({\bf q})$ terms, we may decouple
configurational averaging in (\ref{22}) and write the pair
correlation  function as:
\begin{equation}\label{23}
G_2(|{\bf r}_2-{\bf r}_1|) \simeq  \frac{ {\rm Sp}_\theta e^{-\beta{
\overline{H}}} \overline{ c_{{\bf r}_2}c_{{\bf r}_1}}\Big (1 -
(\theta_{{\bf r}_2}-\theta_{{\bf r}_1})^2/2\Big )} { {\rm Sp}_\theta
e^{-\beta{\overline{H}}}} .
\end{equation}
For the configurationally averaged value of the Hamiltonian ${
\overline{H}}$ one gets:
\begin{eqnarray}\nonumber
\overline{{H}}&=& \frac{1}{2}\sum_{{\bf k}} J|{\bf k}|^2a^2\theta_{\bf k}
\theta_{-{\bf k}} + \sum_{{\bf k}_1,{\bf k}_2}
Ja^2{\bf k}_1.{\bf k}_2\overline{\rho({\bf k}_1+{\bf k}_2)}\theta_{{\bf k}_1} 
\theta_{{\bf k}_2}
\\ \nonumber
&=&
 \frac{1}{2}\sum_{{\bf k}} J|{\bf k}|^2a^2\theta_{\bf k}
\theta_{-{\bf k}} \\
&&\nonumber + \sum_{{\bf k}_1,{\bf k}_2}
Ja^2{\bf k}_1.{\bf k}_2\delta({\bf k}_1+{\bf k}_2)(1-p)\theta_{{\bf k}_1}
 \theta_{{\bf k}_2}
\\ \label{24} &=&
 \frac{1}{2}[1-2(1-p)]\sum_{{\bf k}} J|{\bf k}|^2a^2\theta_{\bf k}
\theta_{-{\bf k}}\nonumber\\
&=&[1-2(1-p)] {H_0}.
\end{eqnarray}
Then the formula for the pair correlation function (\ref{23})
reads:
\begin{eqnarray}\label{25}
&&G_2(|{\bf r}_2-{\bf r}_1|)= {\rm const}\times\nonumber\\
&&\times 
{ {\rm Sp}_\theta
e^{-\beta^\prime{H}_0} \Big (1 -
(\theta_{{\bf r}_2}-\theta_{{\bf r}_1})^2/2\Big )} 
/
{ {\rm Sp}_\theta
e^{-\beta^\prime{H}_0}},
\end{eqnarray}
with
\begin{eqnarray}\label{26}
{\rm const}&=&1-2(1-p), \\ 
\beta^\prime&=&[1-2(1-p)]\beta\label{26a}.
\end{eqnarray}

Formula (\ref{25}) for the pair correlation function of the
{\em diluted} $2d$ $XY$ model has the form of the
spin-\-wave-\-approximated pair correlation function of the
{\em pure} system. It differs only by a constant (\ref{26})
which renormalizes also the temperature $\beta$ (\ref{26a}).
Taking that in the spin-wave approximation the pair correlation
function critical exponent for the pure system equals~\cite{Wegner67}:
 \begin{equation}\label{27}
 \eta^{\rm pure}_{SW}=\frac{1}{2\pi J\beta}
 \end{equation}
 one gets the exponent for the diluted system 
substituting in (\ref{27}) temperature $\beta$ by
$\beta^\prime$ (\ref{26a}):
\begin{equation}\label{28App}
 \eta^{\rm diluted}_{SW}=\frac{1}{2\pi J\beta(1-2(1-p))}.
 \end{equation}

\end{appendix}

\end{document}